\documentstyle[manuscript,osa]{revtex}

\begin{document}
\title{Split-step solitons in long fiber links}
\author{Rodislav Driben and Boris A. Malomed}
\address{Department of Interdisciplinary Studies,\\
Faculty of Engineering, Tel Aviv University, Tel Aviv 69978,\\
Israel}
\maketitle

\begin{center}
{\bf ABSTRACT}
\end{center}

We consider a long fiber-optical link consisting of alternating dispersive
and nonlinear segments, i.e., a {\it split-step model} (SSM), in which the
dispersion and nonlinearity are completely separated. Passage of a soliton
(localized pulse) through one cell of the link is described by an
analytically derived map. Multiple numerical iterations of the map reveal
that, at values of the system's stepsize (cell's size) $L$ comparable to the
pulse's dispersion length $z_{D}$, SSM supports stable propagation of pulses
which almost exactly coincide with fundamental solitons of the corresponding
averaged NLS equation. However, in contrast with the NLS equation, the SSM
soliton is a strong {\it attractor}, i.e., a perturbed soliton rapidly
relaxes to it, emitting some radiation. A pulse whose initial amplitude is
too large splits into two solitons; however, splitting can be suppressed by
appropriately chirping the initial pulse. On the other hand, if the initial
amplitude is too small, the pulse turns into a breather, and, below a
certain threshold, it quickly decays into radiation. If $L$ is essentially
larger than $z_{D}$, the input soliton rapidly rearranges itself into
another soliton, with nearly the same area but essentially smaller energy.
At $L$ still larger, the pulse becomes unstable, with a complex system of 
{\it stability windows} found inside the unstable region. Moving solitons
are generated by lending them a frequency shift, which makes it possible to
consider collisions between solitons. Except for a case when the phase
difference between colliding solitons is $\,_{\sim }^{<}\,0.05\cdot \pi $,
the interaction between them is repulsive. 

\newpage

\section{Introduction}

A great potential offered by {\it dispersion management} schemes for
improvement of data transmission by means of solitons in optical-fiber
networks has been well understood (see, e.g., a special journal issue \cite
{journal} and a collection of articles \cite{book}). Dispersion management
is based on periodic compensation of the accumulated dispersion in a
nonlinear optical fiber by means of additional fiber segments having strong
opposite dispersion \cite{DM}. An objective of the present work is to
introduce and study a related but different scheme (a {\it split-step model}%
, or SSM), in which pieces of a dispersive fiber with negligible
nonlinearity periodically alternate with fiber segments operating close to
the zero-dispersion point. In the latter segments, the dispersion is
negligible, while nonlinearity is dominating. In the limit when both the
dispersive and nonlinear segments are very short, the scheme is nothing else
but the commonly known split-step algorithm for simulations of the nonlinear
Schr\"{o}dinger (NLS) equation, which is usually employed in combination
with the fast Fourier transform solving the equation at the linear step \cite
{NM}; the algorithm can be further modified to adapt it to specific optical
models (see, e.g., Ref. \cite{Korea}). While it is a well established fact
that the split-step algorithm yields very accurate and stable results when
simulating solitons, we here aim to consider a model in which separation
between the dispersive and nonlinear segments is {\em physical}, rather than
numerical, and the lengths of the segments are not small, being comparable
to the soliton's dispersion length $z_{D}$. Understanding possibilities for
the existence of {\it split-step solitons} (robust solitary pulses) and the
study of their dynamics (including interactions between them) in a system of
this type is of a certain fundamental interest. Besides that, the system may
also be of practical interest for two different reasons. Firstly, the
simplest possibility to upgrade a linear system so that to make it possible
to transmit solitons through it is to periodically insert strongly nonlinear
fiber segments or similar elements. Secondly, it is necessary to understand
accuracy limits of the split-step algorithm which can be encountered with
the increase of the stepsize. A detailed formulation of our model is given
below in section 2.

It is relevant to note that SSM qualitatively resembles a ``tandem'' model,
recently introduced by Torner \cite{Torner}, which assumes periodic
alternation of layers with strong dispersion and strong {\em quadratic}
nonlinearity. As it was demonstrated in Ref. \cite{Torner}, this
configuration improves propagation conditions for $\chi ^{(2)}$ solitons in
the temporal domain, providing an efficient compensation of the
group-velocity walkoff.

We stress that the very existence of stable solitary pulses in a long SSM
with a finite (nonsmall) stepsize $L$ is not obvious. Although rigorous
theoretical considerations of the existence problem may be quite difficult,
in section 3 we display direct numerical results, which show that a very
robust pulse can be easily created in SSM. The analysis is based on
numerical iterations of a map which is produced by the analytical solution
of the model within one elementary cell. It is found that, in the case $%
L\,\,_{\sim }^{<}\,\,z_{D}$, launching a pulse with the initial shape taken
as per the fundamental soliton of the usual NLS equation, which is obtained
by averaging SSM, immediately gives rise to a stationary pulse that appears
to remain stable for an indefinitely large number of iterations of the
split-step map. In the cases when the stepsize is essentially larger than
the soliton's dispersion length, the initial pulse in the form of the NLS
soliton emits some radiation and rapidly rearranges itself into another{\it %
\ }soliton with almost the same area but smaller energy. Finally, in the
case $L\gg z_{D}$, the pulses tend to completely decay into radiation.
However, in this case we discover a system of {\em stability windows} on the
axis of the parameter $L$: inside the windows, the soliton is still stable.
The last window exists at a fairly large value of $L$.

In sections 4, we consider the influence of variation of the amplitude of
the initial pulse and of taking it in a chirped form. If the amplitude is
too large, the pulse splits into two moving (separating) solitons. However,
the splitting can be prevented by chirping the pulse (with a right sign of
the chirp). The splitting suppression by chirping gives rise to formation of
a large-amplitude pulse which is quite close to the fundamental soliton.

In section 5 we study collisions between two identical solitons. Although
SSM, unlike its NLS counterpart, is not Galilean-invariant, in simulations
it is quite easy to generate a soliton with a nonzero velocity shift. The
interaction strongly depends on the phase difference $\Delta \phi $ between
the solitons. At nearly all the values of $\Delta \phi $ they repel each
other, while at $\Delta \phi =0$ the interaction is attractive. A
transition from the attraction to repulsion occurs at very
small values of $\Delta \phi $.

Because the above-mentioned results are obtained numerically, they will be
presented in the form of various plots and tables illustrating typical
situations. It appears that this form of the presentation makes it possible
to describe the dynamics of pulses in the system quite adequately.

Lastly, we mention that a similar split-step configuration, including only
two fiber segments, one nonlinear and one dispersive, can be used for
various purposes in the form of a fiber loop. In particular, it has recently
been shown that a loop of this type can operate quite efficiently as a
nonlinear mirror for solitons \cite{Bristol}.

\section{ The model}

The dispersive segment of the fiber is described by the linear
Schr\"{o}dinger equation, 
\begin{equation}
iu_{z}-\frac{1}{2}\beta u_{\tau \tau }=0,  \label{dispersion}
\end{equation}
where $u$, $z$ and $\tau $ are, respectively, the local amplitude of the
electromagnetic waves, propagation distance, and local time ($\tau \equiv
t-z/V$, where $t$ and $V$ are the physical time and mean group velocity of
the wave packet), and $\beta $ is the dispersion coefficient. The main
physical assumption concerning this part of the system is that its
nonlinearity may be neglected (for instance, because the fiber's effective
cross section is large \cite{Agrawal}). Obviously, Eq. (1) can be solved by
means of the Fourier transform in $\tau $, which is used as a part of the
standard split-step numerical algorithm \cite{NM}. In an explicit form, the
solution is (it can also be obtained by means of the Green's function for
the linear Schr\"{o}dinger equation) 
\begin{equation}
u(z,\tau )=\frac{1+i\,{\rm sgn}\beta }{\sqrt{2\pi |\beta |z}}\int_{-\infty
}^{+\infty }u(0,\tau )\exp \left[ -\frac{2i}{\beta z}\left( \tau -\tau
^{\prime }\right) ^{2}d\tau ^{\prime }\right] d\tau ^{\prime }.
\label{Green's}
\end{equation}

In the nonlinear segment, it is assumed that the dispersion may be neglected
(practically, the dispersion can be made quite small in {\it %
dispersion-shifted} fibers \cite{Agrawal}), then one is dealing with the
dispersionless NLS\ equation, 
\begin{equation}
iu_{z}+|u|^{2}u=0,  \label{nonlinearity}
\end{equation}
where the nonlinearity coefficient is normalized to be $1$. We also assume
that the third-order dispersion may be neglected, which is normally true,
unless one is going to consider very narrow pulses \cite{Agrawal}.

The evolution of the field inside the nonlinear segment is described by an
obvious solution, 
\begin{equation}
u(z,\tau )=u(0,\tau )\exp \left[ i|u(0,\tau )|^{2}z\right] ,  \label{phase}
\end{equation}
which is another basic ingredient of the split-step numerical algorithm. We
define the system's elementary cell as an interval between midpoints of two
neighboring nonlinear segments. Thus, the full transformation (map) for the
pulse passing one cell can be represented as a superposition of two
nonlinear phase transformations (\ref{phase}) corresponding to the nonlinear
half-segments at the cell's edges, and the Fourier transform corresponding
to the dispersive segment in the middle of the cell. The transformations (%
\ref{Green's}) and (\ref{phase}) are linked by the condition of continuity
of $u(z,\tau )$ at all the junctions between the segments.

Note that (as well as in the case of the dispersion management) the only
dynamical invariant of the model is the energy 
\begin{equation}
E=\int_{-\infty }^{+\infty }\left| u(\tau )\right| ^{2}d\tau \,.  \label{E}
\end{equation}
A Hamiltonian of the model can be easily defined too, but its value is not
conserved at the jumps between different segments.

Models of this type are usually solved with periodic boundary conditions in $%
\tau $. In order to prevent radiation (in those cases when it is generated)
from re-entering the integration domain due to the periodicity, the linear
equation (\ref{dispersion}) was solved directly by means of numerical
methods in a broad integration interval $\Delta T$, placing {\em absorbers}
at its edges (i.e., we did not actually use the Green's-function transform (%
\ref{Green's})). In fact, the absorbers emulate a real physical effect,
namely, the fact that the energy emitted with radiation is lost for the
pulses.

The present model does not take into regard intrinsic fiber losses,
amplification, and filtering. While all these effects are, of course,
important in the real communication networks \cite{Agrawal}, it is well
known that they may be neglected in the first approximation in the case when
the amplification spacing is smaller than the cell size $L$ (see, e.g., Ref. 
\cite{Taras}, where this issue was considered in detail in terms of the
dispersion management), although they may become important (for instance,
giving rise to a specific instability \cite{Matsumoto}) when the spacing
between the amplifiers and filters coincides with $L$. In any case, the
first objective is to study main features of the soliton dynamics in the
large-stepsize SSM, while the losses, gain, and filtering may be added later.

If the lengths of the dispersive and nonlinear segments are $L_{D}$ and $%
L_{N}$, the cell size is $L=L_{D}+L_{N}$. Note that the length $L_{D}$ can
be rescaled so that to make $|\beta |=1$ in Eq. (\ref{dispersion}). Our
simulations (many iterations of the above-mentioned map) have shown that in
the case of the normal dispersion, $\beta >0$, no quasi-stable soliton can
be found (in complete accord with the absence of the bright solitons in the
NLS equation with the normal dispersion \cite{Agrawal}). Therefore, in what
is following below, we are solely dealing with the case $\beta =-1$.

With regard to the normalizations adopted, an average ({\it distributed})
version of the present SSM takes an obvious form, 
\begin{equation}
iu_{z}+\frac{1}{2(n+1)}u_{\tau \tau }+\frac{n}{n+1}|u|^{2}u=0,  \label{NLS}
\end{equation}
where $n\equiv L_{N}/L_{D}$ is a relative measure of the strength of the
nonlinearity in comparison with the dispersion. Note, however, that Eqs. (%
\ref{dispersion}) and (\ref{nonlinearity}) are separately invariant with
respect to transformations 
\begin{equation}
\tau \rightarrow \tau /\Lambda _{D},z\rightarrow z/\Lambda _{D}^{2},\,{\rm %
and}\,\,u\rightarrow \Lambda _{N}u,z\rightarrow z/\Lambda _{N}^{2}
\label{scaling}
\end{equation}
with two independent arbitrary scaling factors $\Lambda _{N}$ and $\Lambda
_{D}$. This also transforms the length ratio, $n\rightarrow \left( \Lambda
_{D}/\Lambda _{N}\right) ^{2}n$, which may be employed to fix the parameter $%
n$. Throughout the paper, we fix $n\equiv 3$.

In the numerical analysis of SSM, we, will first of all, launch an initial
pulse which coincides with the fundamental-soliton solution to the average
equation (\ref{NLS}), $u_{0}(\tau )=\left( \eta /\sqrt{n}\right) ${\rm sech}$%
\left( \eta \tau \right) $, $\eta $ being an arbitrary parameter that
determines the soliton's amplitude and width. Note that, with fixed $\beta
=-1$ and $n=3$, our model, similarly to the usual NLS equation, is still
invariant against the scale transformation (\ref{scaling}) with $\Lambda
_{N}=\Lambda _{D}$. Using this remaining invariance, we fix $\eta =1$ in the
initial pulse, so that 
\begin{equation}
u_{0}(\tau )=\left( 1/\sqrt{n}\right) {\rm sech\,}\tau .  \label{initial}
\end{equation}
Besides the initial pulses (\ref{initial}) which correspond to the
fundamental-soliton solutions to the NLS equation (\ref{NLS}), we will also
be using initial shapes differing from (\ref{initial}) by taking an
arbitrary amplitude, and also by adding chirp to the pulse.

Thus, after fixing all the scales, there remain the single free parameters
in the model, the stepsize $L$, to which there may be added parameters of
the initial pulse (the chirp and relative amplitude) if it is different from
the fundamental NLS soliton.

\section{The fundamental soliton in the split-step system}

\subsection{A moderately large stepsize}

We start with the case $L=1$, when the stepsize is not essentially different
from the dispersion length $z_{D}\sim (n+1)$, defined for the pulse (\ref
{initial}) as per the averaged equation (\ref{NLS}). Obviously, this case is
already drastically different from the usual version of SSM with a very
small stepsize, which is employed in the numerical algorithms for
simulations of the NLS equation.

In Fig. 1 we display a typical numerical solution for this case with the
above-mentioned fixed value $n=3$. As is seen, on a fairly long propagation
distance comprising $1500$ stepsizes (cells) the input (\ref{initial})
generates a very stable pulse propagating virtually without any distortion.
In the application to optical communications, the physical value of the
soliton's dispersion length is, typically, $\sim 100$ km \cite{Agrawal},
hence, in the present case with $n+1=4$, the interval $z=1500$ corresponds
to the actual propagation distance $\sim 30,000$ km.

Despite the fact that there is no visible degradation of the pulse in Fig.
1, we stress that the existence of the soliton in the model with a finite
stepsize in the rigorous sense is not obvious at all, therefore it is
necessary to search for a very weak decay of the pulse, provided that any
decay does take place. To this end, Fig. 1b shows the evolution of the
pulse's energy, $E=\int_{{\rm slot}}|u(\tau )|^{2}d\tau $ (cf. Eq. (\ref{E}%
)), where the integration is performed over a slot which completely covers
the pulse but is much smaller than the total size of the integration domain,
the latter one being $\Delta T=200$. As is seen from Fig. 1b, after an
initial transient stage, a systematic loss of energy at an extremely small
rate takes place indeed. One may interpret this as very weak emission of
radiation, plausibly due to the fact that the soliton does not exist in the
rigorous sense. An extrapolation of the trend seen in Fig. 1b suggests that
the ``quasi-soliton'' that we are dealing with may persist over an extremely
long propagation distance corresponding to $\sim 5\cdot 10^{6}$ stepsizes.
In the actual simulations the soliton kept its shape intact as long as the
simulations could be run (up to $\sim 5\cdot 10^{4}$ stepsizes).

This and many other runs of the simulations strongly suggest that, in the
case $L\sim 1$, the input in the form of a fundamental soliton of the
corresponding averaged NLS equation gives rise to a very robust stationary
pulse that may be regarded as a fundamental soliton of SSM (even if the
soliton does not exist in the rigorous sense, we may consider it as a
virtually existing object). The next natural step is to consider an input
essentially differing from that given by Eq. (\ref{initial}). There are two
straightforward ways to modify the input: changing its amplitude by an
arbitrary factor $A_{0}$, and/or adding some chirp $b_{0}$ to it \cite
{Sweden}, so that the modified initial pulse is 
\begin{equation}
u_{0}(\tau )=\left( A_{0}/\sqrt{n}\right) \exp (ib_{0}\tau ^{2})\cdot {\rm %
sech\,}\tau .  \label{modified}
\end{equation}

It is well known, from both the variational approximation applied to the
perturbed soliton in the form (\ref{modified}) and from direct numerical
simulations, that in the usual NLS equation the input (\ref{modified}) gives
rise to long-lived internal vibrations of the soliton, which are slowly
damped by radiation losses; eventually, the soliton will assume the static
fundamental shape with a smaller value of the energy \cite{Sweden}. However,
if the perturbation of the fundamental soliton is too strong, it can
completely destroy the pulse. In particular, in the case $b_{0}=0$ it is
known from the exact solution obtained by means of the inverse scattering
transform that the soliton will disappear if $A_{0}\leq A_{{\rm thr}}\equiv
1/2$ \cite{Satsuma} (the variational approximation predicts a higher
destruction threshold, $A_{{\rm thr}}=1/\sqrt{2}$ \cite{Sweden}). When the
chirp is too strong, it also destroys the soliton \cite{Sweden}. The
variational approximation predicts that, in the case $A_{0}=1$, the chirp
kills the soliton if $b_{0}^{2}\geq b_{{\rm thr}}^{2}=1/\pi ^{2}$, but it is
known from numerical simulations that the soliton still survives at somewhat
larger values of the chirp \cite{Sweden}.

Typical results illustrating the evolution of the pulses generated by the
unchirped ($b_{0}=0$) initial shape (\ref{modified}) perturbed by $A_{0}\neq
1$ are displayed in Figs. 2 and 3, in which $A_{0}$ takes values,
respectively, $\sqrt{3}\approx \allowbreak 1.\,\allowbreak 73$, $0.4\sqrt{3}%
\approx 0\allowbreak .\,\allowbreak 69$, and $0.2\sqrt{3}\approx
0\allowbreak .\,\allowbreak 346$. As is seen in Fig. 2a, in the case $%
A_{0}>1 $ the pulse quickly rearranges itself, after several vibrations,
into a quasistationary state, which can be verified to be very close to the
fundamental soliton. The rearrangement is accompanied by a considerable loss
of energy, which is radiated away. In fact, the radiation losses make the
fundamental soliton in SSM an effective {\it attractor}, similar to those in
dissipative systems, although there are no direct losses in the present
model.

The strength of the soliton in SSM as the effective attractor is clearly
seen in comparison with its NLS counterpart. To this end, in Fig. 2c and 2d
we display results of direct simulations of the averaged NLS equation (\ref
{NLS}) with exactly the same initial pulse as in Fig. 2a (in fact, the
integration of the NLS equation was also performed by means of the
split-step algorithm, but with a very small stepsize). A drastic difference
between the relaxation of the deformed soliton in two models is obvious,
despite the fact that the shapes of the fundamental soliton in these models
are virtually identical. A natural explanation to this is the fact that SSM
with $L\sim 1$ is strongly nonintegrable, on the contrary to the integrable
NLS equation, and, as it is well known, irreversible dynamical processes
(such as the separation of a soliton and radiation) are much faster in
nonintegrable systems \cite{review}. The fast relaxation of the perturbed
soliton in SSM has an apparent advantage for the applications to
telecommunications; on the other hand, fast relaxation assumes a burst of
radiation emission. In a more realistic model, including optical filters,
the radiation will be absorbed by the filters.

A qualitatively different result (splitting of the initial pulse into two)
is produced by the multiplication of the input by a still larger factor $%
A_{0}$. This effect will be considered separately in the next section.

The perturbation of the initial pulse by the multiplier $A_{0}$ which is
considerably smaller than $1$ produces a very different effect, transforming
the stationary soliton into a {\it breather} with long-period undamped
internal vibrations (Fig. 3). Finally, if $A_{0}$ is too small, this results
in quick destruction of the pulse (Fig. 4). Thus, a threshold value of $%
A_{0} $ separating the cases when the soliton survives and those when it
perishes is located somewhere between $0.69$ and $0.346$ (recall that $A_{%
{\rm thr}}=1/2$ for the NLS equation). An accurate value of $A_{{\rm thr}}$
can be found in a straightforward way by means of longer simulations.

A different generic type of the perturbation is chirping the input. A
typical example with a strong chirp is shown in Fig. 4. In fact, this is
another manifestation of the remarkable robustness of the soliton in SSM:
the pulse quickly sheds off considerable amounts of radiation, loosing with
it nearly half of its initial energy (Fig. 4b), and again rearranges itself
into a quasistationary state. In Fig. 4c, we compare the shapes of the
central parts of the input and output pulses, along with the chirp
distributions in them (the local chirp shown in Fig. 4c is $\phi _{\tau \tau
}$, where $\phi (z,\tau )$ is the internal phase of the pulse). Note that
the final pulse has the amplitude and width which are simultaneously smaller
than those of the initial pulse (this is possible in view of the
considerable radiative losses). As for the chirp in the final pulse, two
features are noteworthy: its absolute values are much smaller than the chirp 
$2b_{0}=2$ in the initial pulse, and, moreover, the average value of the
chirp across the central part of the output pulse is nearly zero. Thus, we
conclude that SSM quite efficiently {\em suppresses the chirp}, which is not
the case in the usual NLS equation. In fact, because the chirp is a
dynamical variable conjugate to the soliton's width, within the framework of
the variational approximation \cite{Sweden}, the chirp suppression is
closely related to the above-mentioned fast suppression of the pulse's
internal vibrations in SSM (Fig. 2a). Nevertheless, simulations with very
large values of $b_{0}$ (not shown here) demonstrate that a very strong
chirp destroys the pulse, as one would expect from the analogy with the NLS
equation.

\subsection{A very large stepsize}

In the case $L\gg 1$, i.e., when the stepsize is considerably larger than
the soliton's dispersion length, the situation is very different from that
described above. This case will take place if, in order to increase the data
transmission rate, narrower solitons are launched into the fiber link \cite
{journal,book} (note that $z_{D}$ scales inversely proportional to the
soliton's temporal width \cite{Agrawal}).

A drastic difference revealed by the numerical computations in the case $%
L\gg 1$ is that the initial pulse in the form of the NLS fundamental soliton
(\ref{initial}) no longer remains a quasistationary solution of SSM.
Instead, as it is illustrated by Fig. 5 pertaining to $L=10$, the initial
pulse undergoes fast evolution, decreasing its amplitude and getting
broader. Nevertheless, this does not lead to decay of the pulse, and it
keeps more than half of its initial energy (Fig. 5b). The outcome pulse can
be very accurately fitted to the soliton-like {\it ansatz,} 
\begin{equation}
|u|=A\,{\rm sech}\left( \tau /T\right) ,  \label{ansatz}
\end{equation}
with some amplitude $A$ and width $T$, and Fig. 5d shows that the pulse's
chirp (taken at a value of $z$ corresponding to the junction between two
cells) remains very small.

The same ansatz (\ref{ansatz}) provides for a good fit for the outcome
solitons generated by simulations at other large values of $L$ (at which the
solitons are still stable, see below). The values of the fitting parameters
for several large stepsizes $L$, all corresponding to $n=3$ and the initial
pulse in the form (\ref{initial}), are collected in Table 1. In the same
table, we also give values of the numerically measured {\it area} of the
pulse corresponding to the expression (\ref{ansatz}), 
\begin{equation}
S\equiv \int_{-\infty }^{+\infty }\left| u(\tau )\right| d\tau  \label{S}
\end{equation}
(unlike the energy (\ref{E}), the area is not a model's dynamical
invariant). A well-known theorem states that an arbitrary pulse which is
considered as an initial condition to the averaged NLS equation (\ref{NLS})
cannot give rise to a soliton in this equation unless the pulse's area
exceeds a threshold value \cite{Zakharov} $S_{\min }=n^{-1/2}\ln \left( 2+%
\sqrt{3}\right) \approx 1.317\cdot n^{-1/2}$; in particular, the area of the
NLS fundamental soliton (\ref{initial}) is $S_{{\rm sol}}=\pi n^{-1/2}$.

It is clearly seen from Table 1 that the soliton's area remains fairly close
to (slightly larger than) the area of the NLS soliton, $S_{{\rm sol}}=\pi /%
\sqrt{3}\approx \allowbreak 1.\,\allowbreak 81$ (recall we have set $n=3$).
Thus, we infer that launching a fundamental NLS\ soliton into SSM with a
large stepsize $L$ transforms the soliton (unless it is unstable, see below)
into another one which belongs to the same family of the fundamental
solitons, having nearly the same area, but with an essentially smaller
energy. The data from Table 1 suggests that, roughly, the residual energy of
the outcome fundamental soliton decreases $\sim 1/\sqrt{L}$, although there
are considerable deviations from this power law. Such a transformation of
one fundamental soliton into the other is a manifestation of strong
nonintegrability of SSM with large $L$.

It may happen that the outcome pulse obtained at large $L$, i.e., a
fundamental soliton of SSM proper, does have a systematic difference in its
shape from its NLS counterpart. For instance, we tried to fit the data to an
ansatz more sophisticated than (\ref{ansatz}), viz., $|u|=A\,\left[ {\rm sech%
}\left( \tau /T\right) \right] ^{\alpha }$. The result was that $\alpha $
may take values between $1$ and $1.6$, having then the width somewhat larger
than that given in Table 1. However, the available accuracy of the numerical
data does not allow us to make a decisive conclusion that $\alpha >1$
provides for a really better fit than the simplest ansatz (\ref{ansatz}).

Lastly, we stress that it may be quite plausible that SSM does not have any
soliton solution in the rigorous sense. However, in practical terms, it is
quite easy to distinguish between cases when virtually stable solitons
persist on extremely long propagation distances, and those when any pulse
quickly decays, as it is shown in detail in the next subsection.

\subsection{Stability windows}

As it was already shown above, a general trend is that, with the increase of 
$L$, the amplitude of the persisting soliton decreases, and at extremely
large values of $L$ solitons decay into radiation. However, detailed
simulations reveal a very nontrivial feature: there are stability windows on
the axis of the parameter $L$, alternating with regions in which the soliton
is either completely unstable or ``semi-unstable''. The latter means that
the soliton suddenly jumps down to a smaller amplitude, at which it seems to
persist. Increasing $L$ by steps $\Delta L=1$, we have found that the
solitons remain continuously stable at $L\leq 14$; then, the instability or
semi-instability takes place at 
\begin{eqnarray}
15 &\leq &L\leq 17,\,L=19,\,L=21,\,L=23,\,L=25,\,27\leq L\leq 34,  \nonumber
\\
\,37 &\leq &L\leq 50,\,52\leq L\leq 58,\,{\rm and}\,\,L\geq 60\,.
\label{unstable}
\end{eqnarray}
Accordingly, stability windows were found at 
\[
L=18,\,L=20,\,L=22,\,L=24,\,L=26,\,L=36,\,L=51,{\rm \,and}\,L=59\,. 
\]

To illustrate these features, we display several examples in Fig. 6: the
last representative of the continuously stable solitons at $L=14$ (a), the
first semi-unstable one at $L=15$ (b), the first fully unstable case at $%
L=16 $ (c), and the first stability window at $L=18$ (d). We do not show
here solitons belonging to the farthest stability windows, as it is rather
difficult to display them using the same scale as in Fig. 6. However, in
Fig. 7 we separately show the eventual shape of the soliton in the last
stability window ($L=59$).

The above system of the stability windows was found with a rather crude
resolution, $\Delta L=1$; we expect that using a finer resolution would
reveal a much more complex system of windows, and it seems plausible that
their total number is infinitely large, i.e., the window system may have a
fractal structure.

Lastly, we notice that, in the stability windows corresponding to the
largest values of $L$, the soliton does not get completely separated from
radiation, which is illustrated by Fig. 7 (cf. Fig. 5c pertaining to $L=10$,
where there is no visible overlapping with radiation).

A complex system of alternating windows was earlier discovered in a very
different nonlinear model based on the Goldstone (also called $\phi ^{4}$)
equation for a real function $\phi (x,t)$, 
\begin{equation}
\phi _{tt}-\phi _{xx}-\phi +\phi ^{3}=0\,.  \label{phi}
\end{equation}
Numerical simulations of collisions between topological solitons (kinks)
with opposite polarities and opposite velocities $\pm v$ in Eq. (\ref{phi})
had revealed that the collision results in annihilation of the kinks into a
breather (which is subject to subsequent slow radiative decay) if the
velocity is very small, and the collision is quasi-elastic if $v$ is
sufficiently close to the limiting velocity $v_{\max }=1$. Between these two
cases, a system of alternating windows for annihilation and quasi-elastic
collisions was found (see Ref. \cite{Campbell} and references therein). A
semi-quantitative explanation to these findings was based on consideration
of exchange between the kink's kinetic energy and the energy absorbed by an
internal oscillatory degree of freedom which the kink is known to have \cite
{Campbell}. It may happen that the SSM soliton also has an internal degree
of freedom, which may be related to the system of the stability/instability
windows. However, a detailed explanation of this feature should be a subject
of a separate work.

To conclude this section, it may be also relevant to mention that, in the
studies of the dispersion-management models, it has been recently found that
stable propagation of a soliton is possible when the average dispersion
takes a slightly normal value (corresponding to $\beta >0$ in Eq. (\ref
{dispersion})) \cite{DM,Taras,normal}. On the basis of additional
simulations, we have concluded that, in the present model, no (quasi)stable
soliton can be found if the dispersion is normal.

\section{Splitting of a large initial pulse and its suppression by means of
chirping}

Getting back to the ``moderate case'' with $L=1$, we now aim to consider the
evolution of the pulse (\ref{modified}) with large values of the amplitude $%
A_{0}$. In this case, the most interesting observation is spontaneous
splitting of the pulse into two ``splinters'' when $A_{0}$ exceeds a certain
threshold value, which is close to $2$ if $L=1$. A typical example of the
onset of the splitting regime is shown in Fig. 8. In the figure, we keep $%
A_{0}=2.2$, but increase $L$ from $1/3$ to $1$. Note that in the NLS
equation, which corresponds to the limit $L\rightarrow 0$, the unchirped ($%
b=0$) pulse (\ref{modified}) does not split at arbitrarily large values of $%
A_{0}$; instead, it gives rise to a higher-order soliton (breather) \cite
{Satsuma}. In contrast with this, Fig. 8 clearly suggests that, with the
increase of $L$, i.e., with increasing departure from the NLS limit, the SSM
pulse with a sufficiently large amplitude and without initial chirp performs
a transition from a strongly pulsating breather to splitting. The splitting
in the absence of chirp seems to be a characteristic feature of
nonintegrable models, as it is definitely impossible in the integrable NLS
equation. Note that a similar feature was recently observed in simulations
of a different nonintegrable model, viz., the cubic-quintic NLS equation 
\cite{Akhmed}.

It is well known that initial chirp can easily split the pulse even in the
integrable NLS equation (see, e.g., Ref. \cite{Kaup}). Therefore, it is
natural to assume that adding the chirp (with a proper sign) to the initial
pulse which splits in the absence of chirp may balance the trend to the
splitting and, eventually, {\em suppress} it, providing for the generation
of a quasistationary nearly fundamental soliton. Simulations show that this
may indeed be the case, a typical example of which is displayed in Fig. 9.
As is seen in Fig. 9b, more than half of the initial energy of the pulse is
lost with emitted radiation, but one may say that a single soliton is
eventually formed, with a small ``satellite'' attached to it, cf. Fig. 8c.
Note that the average value of the chirp in the central body of the output
pulse (Fig. 9c) is close to zero. Thus, properly selected prechirping of the
large-amplitude pulse not only can prevent its splitting, but also helps to
transform it into a nearly fundamental soliton, rather than leaving it in a
strongly vibrating state (cf. Figs. 9a and 8a,b).

On the other hand, adding the chirp with the opposite sign to the initial
large-amplitude pulse, we observed that the splitting, instead of being
suppressed, is strongly enhanced (not shown here). It is, of course, quite
natural that the effect of chirping crucially depends on its sign.

\section{Collisions between solitons in the split-step system}

Equation (\ref{nonlinearity}) and, hence, SSM as a whole, is {\em not}
Galilean invariant, unlike the classical NLS equation. Nevertheless, a
soliton with a nonzero shift of its (inverse) velocity can be generated by
means of a simple numerical trick \cite{Gershon}: a known solution for the
``quiescent'' soliton should be multiplied, at the initial point $z=0$, by $%
\exp (-i\omega \tau )$ with an arbitrary frequency shift $\omega $, which
can generate a velocity shift via the dispersive part of the model. Thus, we
can easily generate two solitons with opposite velocities and simulate their
collision (in fact, moving solitons generated by the splitting of a
large-amplitude initial pulse were already observed above in Fig. 8).

If the frequency shift $\omega $ is large enough, the collision seems quite
elastic, see a typical example in Fig. 10. At smaller frequency shifts,
however, collisions may give rise to very different results. Below, we
display a set of typical examples obtained with $\omega =\pm 0.05$, while $%
L=1$ and $n=3$ (the same values for which a majority of the above results
have been presented).

The outcome of the collision in the case of small $\omega $ strongly depends
on the phase difference $\Delta \phi $ between the colliding solitons (we
consider only collisions between identical ones). As is well known, the
interaction between solitons of the NLS type is attractive if their phase
difference is zero, and repulsive if $\Delta \phi =\pi $. In accord with
this, we observed that the solitons readily pass through each other -
obviously, due to the attraction - in the case $\Delta \phi =0$ (Fig. 11a),
while in the cases $\Delta \phi =\pi $ and $\Delta \phi =\pi /2$ they
clearly repel each other and avoid mutual passage (see Figs. 11b and 11c).
More detailed simulations have demonstrated that, in fact, the interaction
between the solitons is attractive only if the phase difference between them
is very small, $\,_{\sim}^{<}\;0.05\cdot\pi$.

\section{Conclusion}

We have introduced a model of a long fiber-optical link consisting of
alternating dispersive and nonlinear segments, i.e., a model with a complete
separation of the dispersion and nonlinearity. Losses and gain were assumed
to be compensated, therefore they were not included into the model.

Passage of a soliton (localized pulse) \ through one cell of the system was
described by an analytically derived map. Multiple numerical iterations of
the map reveal that, at values of the system's stepsize $L$ comparable to
the soliton's dispersion length $z_{D}$, the system supports indefinitely
long stable propagation of pulses which almost exactly coincide with
fundamental solitons of the corresponding averaged NLS equation. If the
initial pulse is perturbed, it quickly relaxes into the fundamental state,
shedding off some radiation. The relaxation is much faster than in the
averaged NLS equation, which allows one to consider the soliton in the
present model as an effective attractor. This property has an advantage for
telecommunications. However, an initial pulse whose amplitude is too large
splits into two moving solitons. The splitting could be suppressed by
appropriately chirping the initial pulse, so that a nearly fundamental
soliton appears again. On the other hand, if the amplitude of the input
pulse is too small, it turns into a long-period breather, and, below a
certain threshold, it quickly decays into radiation.

If $L$ is essentially larger than $z_{D}$, an input pulse in the form of an
NLS soliton rapidly rearranges itself into another soliton, whose area
remains nearly the same as that of the NLS fundamental soliton, but the
energy is considerably smaller than that of the input pulse. With further
increase of $L$, the pulse becomes unstable; however, a complex system of 
{\it stability windows} was found inside the unstable region.

Moving solitons could be prepared by initially giving them a frequency
shift, which makes it possible to simulate collisions between solitons. A
result is that, except for a case when the phase difference $\Delta \phi $
between colliding identical solitons is very small, 
the interaction between the solitons is repulsive,
and they avoid passing through each other, unless their relative velocity is
very large.

\newpage

\begin{center}
\begin{tabular}{l|cccccccc}
$L$ & $5$ & $10$ & $14$ & $20$ & $26$ & $36$ & $51$ & $59$ \\ \hline
$A$ & $0.45$ & $0.33$ & $0.28$ & $0.24$ & $0.20$ & $0.16$ & $0.12$ & $0.10$
\\ \hline
$T$ & $1.28$ & $1.88$ & $2.15$ & $2.50$ & $2.98$ & $3.60$ & $5.10$ & $5.90$
\\ \hline
$S$ & $1.83$ & $1.87$ & $1.87$ & $1.85$ & $1.87$ & $1.83$ & $1.84$ & $1.84$%
\end{tabular}

$\allowbreak \allowbreak $
\end{center}

Table 1. Values of the fitting parameters from the ansatz (\ref{ansatz}) for
the outcome soliton in the split-step system with $n=3$ and large values of
the stepsize $L$ (the initial pulse was always taken in the form (\ref
{initial})). Also given in the table are values of the soliton's area (\ref
{S}) (note that, for $n=3$, the area of the NLS soliton is $S_{{\rm sol}%
}=\pi /\sqrt{3}\approx \allowbreak 1.\,\allowbreak 81$). \newpage

\section*{Figure captions}

Fig. 1. The numerical solution of the split-step model with the stepsize $%
L=1 $ and the ratio of the lengths of the nonlinear and dispersion segments $%
n=3$. The initial pulse is the fundamental soliton (\ref{initial}) of the
corresponding averaged NLS equation (\ref{NLS}). Shown are (a) $|u|^{2}$ vs. 
$z$ and $\tau $, and (b) the evolution of the pulse's energy over $1500$
stepsizes.

Fig. 2. The same as in Fig. 1 in the case when the perturbed initial pulse ( 
\ref{modified}) is taken with $b_{0}=0$ and $A_{0}=1.73$; the panels (c) and
(d) additionally show evolution of the same pulse and its energy in the
averaged NLS\ counterpart (\ref{NLS}) of the split-step model.

Fig. 3. The same as in Fig. 1a in the case when the initial pulse is
multiplied by $0.69$ (a) or $0.346$ (b).

Fig. 4. The same as in Fig. 1 in the case when the chirp, $b_{0}=1$, is
added to the initial pulse (\ref{modified}) with $A_{0}=1$. The panel (c)
additionally shows the initial (dotted) and final (solid) shapes of the
central part of the pulse, $|u(\tau )|^{2}$, and the distribution of chirp,
therein.

Fig. 5. The same as in Fig. 1 with $L=10$ and $n=3$. The panel (c)
additionally shows the best fit of the shape of the final pulse $|u(\tau )|$
to the ansatz (\ref{ansatz}), and (d) is the distribution of chirp in the
central part of the output pulse.

Fig. 6. Alternation of stable and unstable pulses at large values of the
stepsize (for $n=3$): (a) $L=14$ (the end of the continuous stability
region); (b) $L=15$ (the first semi-unstable case); (c) $L=16$ (the first
fully unstable case); (d) $L=18$ (the first isolated stability window).

Fig. 7. The eventual shape of the fundamental soliton, $|u(\tau )|$, at $%
L=59 $ (with $n=3$). The dotted curve is the best fit to the ansatz (\ref
{ansatz}).

Fig. 8. Splitting of the pulse generated by the initial shape (\ref{modified}%
) with $b_{0}=0$ and large additional multiplier $A_{0}=2.2$ in the model
with $n=3$ and different values of the stepsize: (a) $L=1/3$; (b) $L=1/2$;
(c) $L=1$.

Fig. 9. Suppression of the pulse splitting shown in Fig. 8 by prechirping
(with $b_{0}=1$) the initial shape (\ref{modified}): (a) $|u|^{2}$ vs. $z$
and $\tau $; (b) the evolution of the pulse's energy; (c) the distribution
of the chirp in the output field (taken at $z=300$),

Fig. 10. A quasi-elastic collision of two identical solitons in the
split-step model with $L=1$, $n=3$, and the phase difference $\Delta \phi =0$%
. The solitons were set in motion by giving them large initial frequency
shifts $\omega =\pm 0.76$.

Fig. 11. Collisions between two identical solitons that were ``pushed'' by
small initial frequency shifts $\omega =\pm 0.05$ in the case $L=1$ and $n=3$%
. The phase difference between the solitons is $\Delta \phi =0$ (a), $\Delta
\phi =\pi $ (b), and $\Delta \phi =\pi /2$ (c). Because in the case of the
small relative velocity it is difficult to show collisions by means of
three-dimensional plots similar to that shown in Fig. 10, we here instead
display contour plots, that illustrate the collisions process quite clearly.

\end{document}